\documentclass[sigconf, nonacm]{acmart}
\usepackage{subcaption}
\usepackage{graphicx}

\newcommand\holodoi{10.61981/ZFSH2303}
\newcommand\holopages{29-33}
\newcommand\holovolume{1}
\newcommand\holoissue{1}
\newcommand\holoyear{2023}
\newcommand\holoauthors{\authors}
\newcommand\holotitle{\shorttitle} 
\newcommand\holoavailabilityurl{Text is provided}
\newcommand\holopagestyle{plain} 

\begin{document}
\title{Flight Patterns for Swarms of Drones}

\author{Shuqin Zhu, Shahram Ghandeharizadeh}

\affiliation{%
\institution{Computer Science Department, University of Southern California}
  \streetaddress{Computer Science Department}
  \city{Los Angeles}
  \state{California}
  \country{USA}
  \postcode{90089}
}
\email{{shuqinzh,shahram}@usc.edu}

\begin{abstract}
We present flight patterns for a collision-free passage of swarms of drones through one or more openings.  The narrow openings provide drones with access to an infrastructure component such as charging stations to charge their depleted batteries and hangars for storage.  The flight patterns are a staging area (queues) that match the rate at which an infrastructure component and its openings process drones.  They prevent collisions and may implement different policies that control the order in which drones pass through an opening.  We illustrate the flight patterns with a 3D display that uses drones configured with light sources to illuminate shapes.


\end{abstract}

\maketitle

\pagestyle{\holopagestyle}
\begingroup\small\noindent\raggedright\textbf{Holodecks Reference Format:}\\
\holoauthors. \holotitle. Holodecks, \holovolume(\holoissue): \holopages, \holoyear.\\
\href{https://doi.org/\holodoi}{doi:\holodoi}
\endgroup
\begingroup
\renewcommand\thefootnote{}\footnote{\noindent
This work is licensed under the Creative Commons BY-NC-ND 4.0 International License. Visit \url{https://creativecommons.org/licenses/by-nc-nd/4.0/} to view a copy of this license. For any use beyond those covered by this license, obtain permission by emailing \href{mailto:info@holodecks.quest}{info@holodecks.quest}. Copyright is held by the owner/author(s). Publication rights licensed to the Holodecks Foundation. \\
\raggedright Proceedings of the Holodecks Foundation, Vol. \holovolume, No. \holoissue.\\
\href{https://doi.org/\holodoi}{doi:\holodoi} \\
}\addtocounter{footnote}{-1}\endgroup

\ifdefempty{\holoavailabilityurl}{}{
\vspace{.3cm}
\begingroup\small\noindent\raggedright\textbf{Holodecks Artifact Availability:}\\
A circular flight pattern in a horizontal, vertical, and diagonal alignments are shown at \url{https://youtu.be/oT5RR8RPl0I}, \url{https://youtu.be/TQM4hMBwLHM}, and \url{https://youtu.be/NNlWn9VW894}, respectively.
The speed of the drones is 1 meter/second.
\endgroup
}

\section{Introduction}\label{sec:intro}
Unmanned Aerial Vehicles (UAVs), drones, are used in diverse applications such as holodecks~\cite{shahram2021,shahram2022}, haptic interactions~\cite{BeyongTheForce,flshaptic23,flyable}, package delivery~\cite{delivery2021,dronetalk2022,dronescheduling2022}, information gathering~\cite{3dpathplanning2014}, search for and track targets~\cite{routeplanning2012}, and disaster relief and rescue~\cite{disaster2019,fire2018}.
The architecture of a system may require a swarm of drones to pass through one or more openings to access an infrastructure component such as charging (refueling) stations, hangars for permanent storage of drones among others.
This paper presents flight patterns for collision-free passage of drones through an opening.
A flight pattern is a temporary staging area for the drones before they pass through an opening.
It may organize drones in a queue, allowing one drone to arrive at the opening at a time and gain entry to the infrastructure.

To simplify discussion and without loss of generality, we focus on a 3D display using the Flying Light Specks, FLSs~\cite{shahram2021}.
An FLS is a miniature sized drone configured with light sources.
It has its own processor, storage, and networking card.  A swarm of FLSs cooperate to illuminate complex 2D and 3D shapes and animated sequences~\cite{shahram2021,shahram2022}.
An illumination may require a large swarm consisting of millions and billions of FLSs~\cite{shahram2021}.
Once the display of an illumination ends, FLSs must fly back to their hangars for storage.  
Similarly, our current FLSs are battery powered with a fixed flight time on a fully charged battery.
They must fly to a charging station to charge their battery.

Today's display architectures~\cite{shahram2022,dv2023} provide narrow openings for FLSs to access the hangers, charging stations, or both.
A challenge is how to provide FLSs with a collision free passage through these opening.
To illustrate, Figure~\ref{fig:dronevision} shows the architecture of the Dronevision (DV) with glass panes on its side to protect a user from rogue FLSs~\cite{dv2023}.
The top panel has an opening on the side that enables FLSs to fly through to access the charging stations.
The rods on the side are silos that enable the FLSs to be deposited to the hangars at the base of the display.
An advantage of this architecture is that it minimizes the likelihood of failed FLSs from falling on and damaging those FLSs in transit to charge their batteries.


A narrow opening raises the following challenge:  
How to provide collision-free passage to a large number of FLSs trying to fly through one or more of these openings at the same time?
``Large" because hundreds of FLSs may want to charge their battery at the same time, see discussion of STAG~\cite{shahram2022} in Section~\ref{sec:related}.
Similarly, when a display is shut-down, a large number of FLSs will fly back to be stored in a hanager at the same time.  

A key requirement is for an FLS to arrive at its destination as fast as possible, minimizing its transit time.
This enables an FLS to arrive at a charging station with the most flight time remaining, reducing the time the FLS spends charging its battery\footnote{In turn, this frees the charging station for use by other FLSs, reducing the total number of charging stations.}.
In addition, we want to give a higher priority to those FLSs with a lower flight time.
This prevents FLSs from failing by exhausting their battery life time prior to arriving at a charging station in a timely manner.

One approach to satisfy these requirements is to introduce flight patterns.
A flight pattern organizes FLSs in queues to fly through an openings one at a time.  This approach has several advantages.  First, the likelihood of one or more FLSs colliding is reduced.  Second, once admitted in a queue, FLSs may cooperate and re-order themselves to implement different policies. 
An example policy may be to minimize the number of FLSs that fail due to exhausting their battery's remaining flight time.
A mechanism to implement this policy may be to allow the FLS with the least remaining flight time to move to the head of the queue.  

The {\bf contributions} of this paper is the concept of flight patterns for a swarm of drones with a focus on FLSs, see Section~\ref{sec:patterns}.
We describe the related work in Section~\ref{sec:related}.
Brief future work is detailed in Section~\ref{sec:future}.

\section{Flight Patterns}\label{sec:patterns}

We assume an infrastructure such as a hangar or a charging station has an admission rate of $\lambda$ FLSs per unit of time.  An opening is a component of the infrastructure.  We assume the infrastructure is accessible by one opening.
Extensions to multiple opening are described in Section~\ref{sec:multiple}.
With this assumption, the opening may admit one FLS every $\frac{1}{\lambda}$ time units.  To illustrate, with $\lambda$=10 FLSs per second, the opening allows one FLS to pass through every 100 milliseconds.

A flight pattern consists of a fixed number of slots $M$, one per FLS.  The pattern terminates at the opening for an infrastructure component, e.g., a charging station or a hanger.  The slots are interleaved $\frac{1}{\lambda}$ time units apart.  This in combination with the distance between two slots of the pattern dictate the speed of each FLS.

Flight patterns may either be geometric or non-geometric.
A geometric pattern may be a circle, an ellipsoid, a rectangle, etc., see Figure~\ref{fig:geometric_shapes}.
See \url{https://youtu.be/oT5RR8RPl0I} for a circular flight pattern in a horizontal alignment.
A non-geometric pattern may be a zig-zag line such as those found in a busy airport or a popular theme park, see Figure~\ref{fig:line}.

A pattern may be organized in a 2D or a 3D hierarchy.
A 2D hierarchy consists of N patterns, say an ellipsoid, with pattern $i$ containing pattern $i-1$. Figure~\ref{fig:2D_hierarchical} shows a horizontal pattern.  Other alignments are possible as discussed in Section~\ref{sec:future}.

A 3D hierarchy stacks the N patterns atop of one another.  The participating patterns may be either homogeneous or heterogeneous.  A homogeneous hierarchy consists of the same pattern with either identical or different settings.  For example, circles with either the same or different diameters, see Figure~\ref{fig:hierarchical_pattern}. 
A heterogeneous hierarchy consists of different patterns.  Figure~\ref{fig:3D_heterogeneous} shows an example consisting of ellipsoid, rectangle, and zig-zag lines.  


A centralized Orchestrator~\cite{shahram2022} may manage the slots of a flight pattern.
An FLS that wants to enter an opening contacts the Orchestrator and the Orchestrator assigns the last available slot to the FLS.
The coordinates of this slot depends on the number of queued FLSs.
If there are no queued FLSs then its coordinates are the opening.
Otherwise, it is the coordinates of the next empty slot in a queue.
Figures~\ref{fig:geometric_shapes}-\ref{fig:3D_heterogeneous} show the worst case scenario where M-1 slots of a pattern are occupied by an FLS. 
FLSs in a queue may communicate and switch positions to implement a policy, e.g., move the FLS with the least amount of remaining flight time to the head of the queue.

\subsection{Multiple Openings}\label{sec:multiple}
With multiple openings, the relationship between an opening and the infrastructure may either be exclusive, shared, or hybrid.
With exclusive, the infrastructure is partitioned across the $\Theta$ openings, each with its own admission rate, \{$\lambda_1$, $\lambda_2$, ..., $\lambda_\Theta$\}.  A challenge is how to distribute the FLSs across the partitioned infrastructure instances and their openings. 

Shared allows an opening to provide access to the infrastructure in its entirety.  We assume the total consumption rate of the openings equals $\lambda$,
$\sum_{i=1}^{\Theta}\lambda_i=\lambda$.  A challenge is how to distribute the FLSs across the $\Theta$ openings.  
When every opening has a queue of FLSs, a simple technique may assign an FLS to the opening with the shortest queue 
while considering factors such as distance, the remaining battery flight time of the FLS, and the consumption rate of an opening $\lambda_i$.

With hybrid, the infrastructure is partitioned such that one or more partitions may have multiple openings while others have one opening.
It must address the challenges faced by both exclusive and shared for the different openings.

\section{Related Work}\label{sec:related}
To the best of our knowledge, the concept of collision-free flight patterns for swarms of drones is novel and not described elsewhere. 
The concept does exist in aviation with air-planes following a pattern dictated by a control tower to land at busy airports~\cite{AirTraffic}.
The control tower is equivalent to a centralized Orchestrator that assigns slots to drones to provide a collision-free landing.  


There is a signficant body of work on scheduling drones for communication by extending the radio range of devices, delivery of packages in an order that maximizing profits, among others.
See~\cite{schedulingsurvey2022} for a survey of these studies.
A study may use optimization techniques to schedule the way-points visited by a drone.
None present flight patterns for a swarm of drones.
Some of the scheduling techniques motivate the flight patterns of this study.  For example, STAG~\cite{shahram2022} is a scheduling technique that staggers the remaining flight time of FLSs to prevent all FLSs from exhausting their flight time at the same time.
It switches the place of a fully charged FLS with one that has almost depleted its battery every $S$ interval of time.
This staggering interval is dictated by the time for an FLS with an almost depleted battery to arrive at a charging station.

With a large number of FLSs, STAG constructs $h$ flocks, resulting in $h$ FLSs in transit from an illumination to a charging station every $S$ interval of time.
In~\cite{shahram2022}, we analyze an illuminated rose consisting of 65K FLSs.
Each FLS has a 5 minute flight time on a fully charged battery and requires 10 minutes to charge its depleted battery fully.
The rose requires $h$=218 FLSs in transit with a staggering interval of 1.36 seconds.
This is a large number of FLSs traveling to a charging station.
No collision-free technique is presented in~\cite{shahram2022}, motivating the flight patters presented in this paper.

A large number of studies present techniques to either avoid~\cite{Schildbach2016, Tomlin1998, Margellos2013} or detect and prevent~\cite{
Williamson1989, Chen2017, Niwa2017, Gageik2015, Zheng2017} collisions.
These studies may use optimization techniques~\cite{Zhang2021Optimization, Han2022Grid, Margellos2011}, search and planning techniques~\cite{Sun2020DenseAPF, Sun2017APF, Jyoti2021Rogue}, reinforcement learning~\cite{Kaufmann2023ChampionLevel, chen2023asynchronous, Hsu2022}, and nature inspired techniques~\cite{ZHOU2021bio-inspired, Huang2019}.
While they do not present flight patterns, one or more of these techniques may be used to control how drones fly to gain access to their assigned slot in a pattern.
Similarly, they can be used to control how two or more FLSs may switch positions in a queue.

\section{Conclusions and Future Research}\label{sec:future}

\begin{figure}
\centering
\includegraphics[width=0.7\columnwidth]{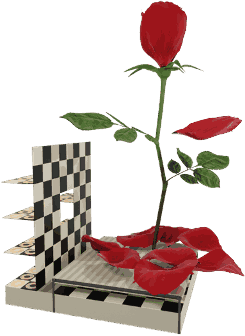}
\caption{A Dronevision (DV) with charging stations behind a wall at the back of the display.  FLSs with depleted batteries fly through the narrow vertical opening to land on a charging station and charge their batteries.}
\label{fig:dv2}
\end{figure}

Flight patterns provide a collision free passage through one or more opening for swarms of drones.
These openings may provide access to a charging station or a hangar for storage of the drones.  The flight patterns are queues that require drones to travel at a certain speed that matches the consumption rate of the infrastructure corresponding to the opening, e.g., hangar, charging station, etc.  These patterns were presented in a horizontal setting for a narrow opening at the top of a display.
However, they may be formed in any alignment.
For example, a vertical formation may be appropriate for the Dronevision of Figure~\ref{fig:dv2}.
The concept of flight patterns raises many interesting research topics.  We present a few in the rest of this section. 

Drones may be assigned to the slots using either a centralized or a decentralized algorithm.  With both, drones should be able to communicate to implement gossip based protocols.
A drone in a queue may communicate with the drones in its radio range, e.g., those drones behind, ahead, above, and below it. 
A message may contain its neighbor's metadata, e.g., remaining flight time.
This information enables two or more drones to implement different policies for managing their ordering in the queue.
Details of how drones change position in a collision free manner are also a future research direction. 

It is important to gain insight into the alternative flight patterns and their tradeoffs.
This is a multi-faceted topic.
It includes the characteristics and limitations of drones as a dimension.
For example, a limitation of FLSs is that they fail~\cite{shahram2021}.
There are different forms of failures including those that cause the FLS to stop flying all together, dropping to the floor of the display.
It is important to understand which flight patterns are more tolerant of FLS failures.  In specific, the failure of an FLS and how it may fall on the FLSs below it is an open research topic.

\begin{figure}
\centering
\includegraphics[width=0.9\columnwidth]{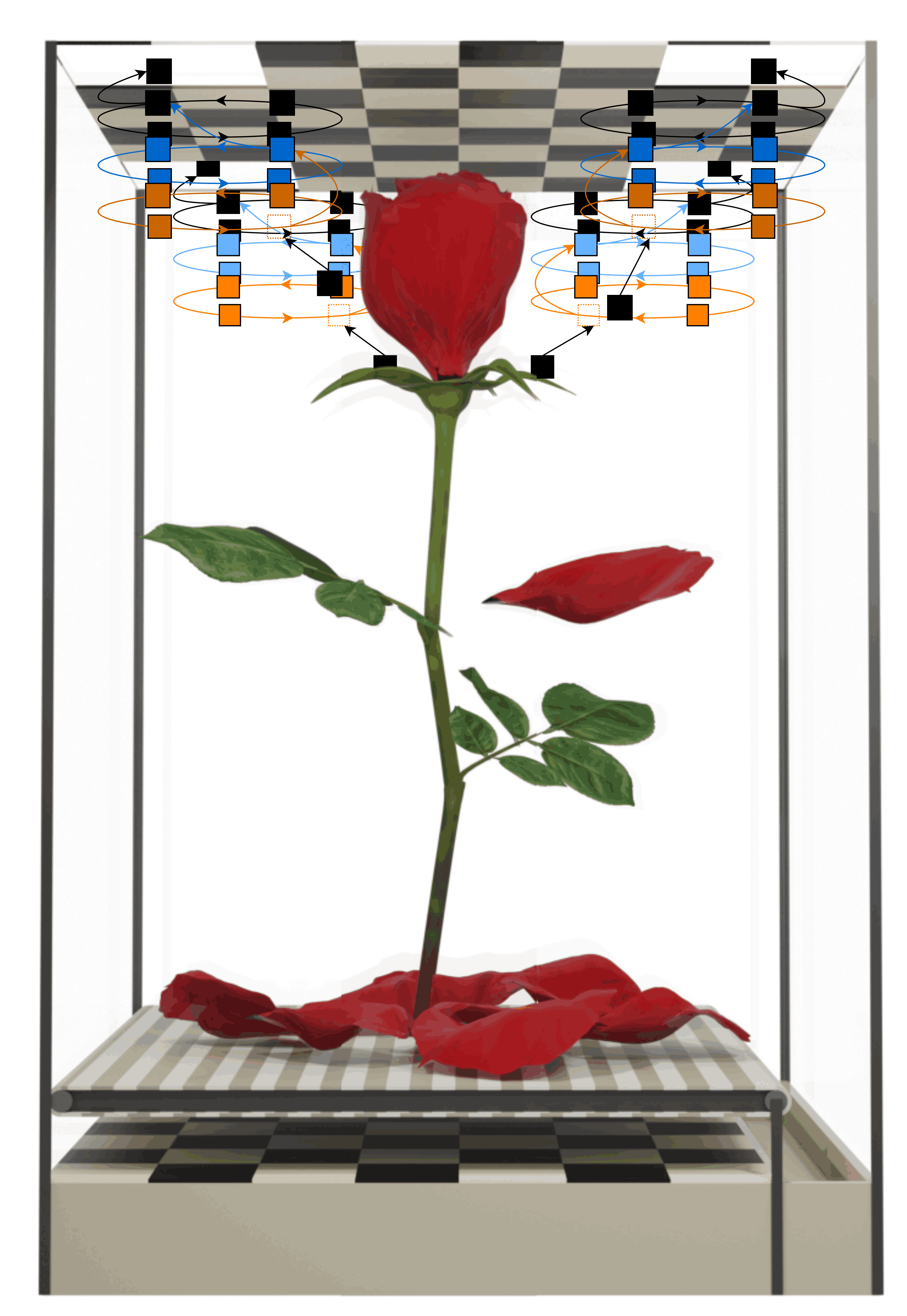}
\caption{Charged FLSs join the bottom of the rose illumination.  They move up in an orderly manner.  Those with almost depleted batteries at the top of the illumination.  These FLSs are scheduled in flight patterns for one of the four openings that provide them with access to the charging coils.}
\label{fig:rosecharge}
\end{figure}

With FLSs, a flight pattern may synergize with an illumination.  
For example, in Figure~\ref{fig:rosecharge}, FLSs with fully charged batteries may join the bottom of an illumination.
They move up to the top of the display based on a regular schedule.
Their battery is almost depleted once they arrive at the very top, causing them to join a flight pattern at one of the entries to access a charging station.
A key question is what happens to the FLSs illuminating the falling petal.
A possible answer is to have a subset of the FLSs move up the illumination more rapidly to illuminate the falling petal.
An investigation of this question and the overall approach is a future research direction.

\balance

\begin{acks}
This research was supported in part by the NSF grant IIS-2232382.
We thank Hamed Alimohammadzadeh for the drawings of the Dronevision presented in this paper and Rohit Bernard for the video recordings of circular Crazyflie flight patterns with different alignments.
\end{acks}


\bibliographystyle{ACM-Reference-Format}
\bibliography{sample}

\end{document}